\documentclass[prd,twocolumn,preprintnumbers,superscriptaddress,nofootinbib,showpacs]{revtex4}
\usepackage[a4paper, hdivide={1.91cm,,1.165cm}, vdivide={1.83cm,,2.6cm}]{geometry}

\usepackage{amstext,amssymb}
\usepackage{amsmath}
\usepackage{graphicx}
\usepackage[hyperfootnotes=false]{hyperref}
\usepackage{xspace}
\usepackage{color}
\usepackage{units}
\usepackage{slashed} % feynman slash notation via \slashed{p}

\newcommand{\dd}{\mathrm{d}}

\def \L {\mathcal{L}} %Lagrangian density

\def \vec#1{{\boldsymbol{#1}}}
\newcommand{\hc}{\ensuremath{\text{h.c.}}}

\begin{document}
%%%%%%%%%%%%%%%%%%%%%%%%%%%%%%%%%

\title{Minimal Left--Right Symmetric Dark Matter}

\preprint{ULB-TH/15-10}

\author{Julian \surname{Heeck}}
\email{julian.heeck@ulb.ac.be}
\affiliation{Service de Physique Th\'eorique, Universit\'e Libre de Bruxelles, Boulevard du Triomphe, CP225, 1050 Brussels, Belgium}

\author{Sudhanwa \surname{Patra}}
\email{sudhanwa@mpi-hd.mpg.de}
\affiliation{Max-Planck-Institut f\"ur Kernphysik, Saupfercheckweg 1, 69117 Heidelberg, Germany}
\affiliation{Center of Excellence in Theoretical and Mathematical Sciences,
Siksha 'O' Anusandhan University, Bhubaneswar-751030, India}

%%%%%%%%%%%%%%%%%%%%%%%%%%%%%%%%%%
%%%%%%%%%%%%%%%%%%%%%%%%%%%%%%%%%%
\begin{abstract}

We show that left--right symmetric models can easily accommodate stable TeV-scale dark matter particles without the need for an ad-hoc stabilizing symmetry. The stability of a newly introduced multiplet arises either accidentally as in the Minimal Dark Matter framework or comes courtesy of the remaining unbroken $\mathbb{Z}_2$ subgroup of $B-L$. Only one new parameter is introduced: the mass of the new multiplet. As minimal examples we study left--right fermion triplets and quintuplets and show that they can form viable two-component dark matter.
This approach is in particular valid for $SU(2)\times SU(2)\times U(1)$ models that explain the recent diboson excess at ATLAS in terms of a new charged gauge boson of mass 2\,TeV.

\end{abstract}

\pacs{12.60.Cn, 95.35.+d}
%12.60.Cn 	Extensions of electroweak gauge sector
%95.35.+d 	Dark matter

%%%%%%%%%%%%%%%%%%%%%%%%%%%%%%%%%
%%%%%%%%%%%%%%%%%%%%%%%%%%%%%%%%%
\maketitle

%%%%%%%%%%%

\section{Introduction}

The discovery of a scalar boson at the LHC has fortified the credibility of the Standard Model (SM) that accounts for the fundamental interactions up to current accelerator energies. But its inadequacy in explaining non-zero neutrino masses, dark matter (DM), and the origin of the matter--antimatter asymmetry of the universe, compels us to extend its horizons.

In this regard, left--right (LR) symmetric models based on the gauge group $SU(2)_L \times SU(2)_R \times U(1)_{B-L}$~\cite{Mohapatra:1974gc, Pati:1974yy, Senjanovic:1975rk,Senjanovic:1978ev,Mohapatra:1979ia,Mohapatra:1980yp} seem appealing, as they provide a clear description of maximal parity violation and pave a path for naturally light neutrino masses.
If the scale associated with the breaking of $SU(2)_R\times U(1)_{B-L}$ occurs at a few TeV, it leads to interesting collider signatures,
neutrinoless double beta decay and lepton flavor violation.
As of now, however, the mystery of DM still remains to be addressed within LR models. 
Lowering the mass of one of the right-handed neutrinos to the keV scale can potentially make it a valid long-lived warm DM 
candidate~\cite{Bezrukov:2009th,Nemevsek:2012cd} -- with interesting signatures in neutrino-mass searches~\cite{Barry:2014ika} 
-- but the approach is far from natural in LR models and requires a delicate production mechanism in the early universe.

In this letter, we explore the possibilities of having stable TeV-scale DM in left--right models. This is easily possible 
by introducing new multiplets that are either accidentally stable~\cite{Cirelli:2005uq} or stabilized by the remaining discrete symmetry $U(1)_{B-L}\to \mathbb{Z}_2$~\cite{Heeck:2015pia}. We will focus on fermionic triplets and quintuplets and show that they can indeed account for the dark matter abundance we observe.

\section{Left--right models}

Under the LR gauge group $SU(2)_L\times SU(2)_R\times U(1)_{B-L}$ -- omitting the $SU(3)_C$ factor for simplicity -- 
the usual fermion content of the SM, plus right-handed neutrinos~$\nu_R$, falls into the following representations:
\begin{align}
 \ell_L &\sim (\vec{2},\vec{1},-1)\,, &
 q_L &\sim (\vec{2},\vec{1}, 1/3)\,,\\
\ell_R  &\sim (\vec{1},\vec{2},-1)\,, &
 q_R &\sim (\vec{1},\vec{2}, 1/3)\,.
\end{align}
To break the symmetry spontaneously down to $U(1)_\text{EM}$ and provide fermion masses, one usually introduces a bi-doublet $H$ and two triplet scalars $\chi_{L,R}$
\begin{align}
 H  \sim (\vec{2},\overline{\vec{2}}, 0) \,, &&
\chi_L \sim (\vec{3},\vec{1},-2)\,,&&
\chi_R \sim (\vec{1},\vec{3},-2)\,.
\label{eq:LRfermions}
\end{align}
A non-zero vacuum expectation value of the neutral component of $\chi_R$, $\langle \chi_R^0 \rangle = v_R/\sqrt{2}$, breaks $SU(2)_R \times U(1)_{B-L}\to U(1)_Y$ at a scale above TeV to generate large Majorana masses 
for the right-handed neutrinos -- leading to seesaw neutrino masses for the active neutrinos -- and 
masses for the new gauge bosons $W_R$ and $Z_R$:
\begin{align}
M_{W_R} \simeq \frac{g_R}{\sqrt{2}} v_R \,, &&
M_{Z_R} \simeq \frac{g_R c_W}{\sqrt{\cos 2\theta_W}} v_R\,,
\end{align}
further modified due to mixing with the left-handed gauge bosons induced by $\langle H\rangle$.
While one typically assumes $g_R = g_L$ due to an additional LR exchange symmetry (parity or charge conjugation), this depends on the full breaking scheme, and so does the ratio $M_{Z_R}/M_{W_R}\simeq 1.7$~\cite{Deppisch:2014qpa,Heikinheimo:2014tba,Deppisch:2014zta,Patra:2015bga}.
The generator of electric charge is given by 
\begin{align}
Q = T^3_L + T^3_R + \tfrac12 (B-L) \,,
\end{align}
where $T^3_X$ denotes the diagonal generator of $SU(2)_X$.

With the quantum numbers from above we see that $U(1)_{B-L}$ is actually broken down to a non-trivial $\mathbb{Z}_2$ 
subgroup by $\langle \chi_R^0 \rangle$, under which all fermions are odd and all bosons are even.\footnote{Actually, one obtains a $\mathbb{Z}_6\cong  \mathbb{Z}_2\times \mathbb{Z}_3$ subgroup, but the $\mathbb{Z}_3$ part is related to baryon number and not of interest here.} Introducing a new fermion (scalar) multiplet with even (odd) $B-L$ charge hence automatically stabilizes the lightest component~\cite{Heeck:2015pia}. 
Another option is to introduce multiplets that are stable only 
at the renormalizable level because it is not possible to write down dimension-4 operators that lead to decay, an idea known as Minimal Dark Matter (MDM)~\cite{Cirelli:2005uq}.

Let us list the simplest possibilities for stable particles in LR models. The chiral fermion multiplets
\begin{align}
 \phi_L \sim (\vec{2 n +1},\vec{1},0)\,, &&
 \phi_R \sim (\vec{1},\vec{2 n +1},0)\,,
\label{eq:fermion_multiplets}
\end{align}
for $n\in \mathbb{N}$, share a common ``Majorana'' mass $M$ due to LR exchange symmetry, which is split by radiative corrections. The left-handed multiplet is simply a MDM candidate~\cite{Cirelli:2005uq,Cirelli:2009uv}, 
while $\phi_R$ can have a different phenomenology depending on the $W_R$ and $Z_R$ masses, to be discussed below.
Another option is a chiral bi-multiplet $(\vec{n},\vec{n},0)$, which 
again allows for a Majorana mass. In particular, $(\vec{2 n +1},\vec{2 n +1},0)$ bi-multiplets contain neutral components without hypercharge and are thus potentially safe from the stringent bounds of direct-detection experiments. 
Putting \emph{real scalars} into the representations of Eq.~\eqref{eq:fermion_multiplets} will not make them absolutely stable 
for all $n$, but only for large enough $n$ in the MDM spirit. For example, $n=1$ would allow 
for a $\eta H^2$ coupling of the triplet scalar $\eta$, $n=2$ for a $\eta \Delta^2$ coupling, so $n> 2$ is required. Higher-dimensional operators of the form $\eta H^4$ or $\eta^3 H^2$, even if suppressed by the Planck scale, can however make the scalar $\eta$ unstable even for $n> 2$, similar to the MDM case~\cite{Cirelli:2005uq,DiLuzio:2015oha}.

For Dirac/complex fields we can write down many more possibilities. For example, a scalar stabilized 
by the $\mathbb{Z}_2$ can be obtained from the multiplets
\begin{align}
 \eta_L \sim (\vec{2},\vec{1},-1)\,, &&
 \eta_R \sim (\vec{1},\vec{2},-1)\,,
\end{align}
which have many non-trivial couplings and a potentially interesting phenomenology. One can extend this to
\begin{align}
 \eta_L \sim (\vec{2 k},\vec{1},2m+1)\,, &&
 \eta_R \sim (\vec{1},\vec{2 k},2m+1)\,,
\end{align}
ensuring that it contains a neutral component.

A thorough study of all these candidates has to be postponed to a longer article~\cite{preparation}. Below we will study only the simplest possibility, fermion triplets and quintuplets.\footnote{Note that the embedding of the triplet into LR models makes it exactly stable. This is not the case in MDM, unless one imposes $B-L$ as a symmetry~\cite{Cirelli:2014dsa}.} 
We stress that the above approach works for quite general $SU(2)_1\times SU(2)_2\times U(1)'$ models, not necessarily LR symmetric. In those cases, the masses of the left- and right-handed multiplets need not be the same, nor does one need to introduce both. Since the gauge couplings are typically also unrestricted in more general models, there is a huge parameter space to be explored.
We focus on the most restrictive case of LR symmetry in order to reduce the number of free parameters.

\section{Minimal model}

Let us take a look at the LR fermion multiplets from Eq.~\eqref{eq:fermion_multiplets}.
Seeing as it only introduces one additional parameter, the common mass $M$, it can be considered the minimal 
LR DM. For each multiplet $\phi_X$ one can write down a Majorana mass term, and due to the LR 
exchange symmetry (parity) $\phi_L\leftrightarrow\phi_R$, both multiplets are degenerate: 
\begin{align}
\L_\phi  = \sum_{X=L,R}\left[\ i \overline{\phi}_X \slashed{D} P_X \phi_X - \frac{M}{2} 
\left(\overline{\phi}_X^c P_X \phi_X + \hc\right) \right] .
\label{eq:MFDM}
\end{align}
Here, $P_{R,L} \equiv (1\pm \gamma_5)/2$ are the usual chiral projection operators.
The charged components of each multiplet $\phi_X$ form massive Dirac fermions $\Psi^m_X \equiv \phi^m_X +(-1)^m (\phi^{-m}_X)^c$, $m=1,\dots,n$ -- 
where $\Psi^m_X$ has electric charge $m$ -- while the neutral ones are Majorana $\Psi^0_X \equiv \phi^0_X + 
(\phi^0_X)^c$. 
(Note that one obtains two stable particles here, one from each $SU(2)$. In principle only the lightest of the two is absolutely stable -- due to the $\mathbb{Z}_2^{B-L}$ -- but since mixing terms only arise at the non-renormalizable level, e.g.~$\phi_L H H \phi_R/\Lambda$, we accidentally end up with two-component DM.)
The gauge couplings of the mass eigenstates are
\begin{align}
\begin{split}
\L_\phi &\supset \sum_{X=L,R}\left[ g_X \sum_{m=1}^n \left(  m \overline{\Psi}^m_X \slashed{W}_{X}^3 \Psi^m_X  \right) \right. \\
&\quad \left.+ \frac{g_X}{\sqrt{2}} \left( \sum_{m=0}^{n-1} c_{n,m} \overline{\Psi}_X^{m+1} \slashed{W}^+_X \Psi^m_X 
+\hc\right) \right],
\end{split}
\end{align}
with $c_{n,m} \equiv \sqrt{(n+m+1)(n-m)}$. 
As always, the gauge eigenstates $W_{L,R}$ mix into the mass eigenstates $W_{1,2}^\pm$, $Z_{1,2}$, and $\gamma$~\cite{Duka:1999uc}.

\subsection{Mass splitting}

\begin{figure}[t]
\includegraphics[width=0.45\textwidth]{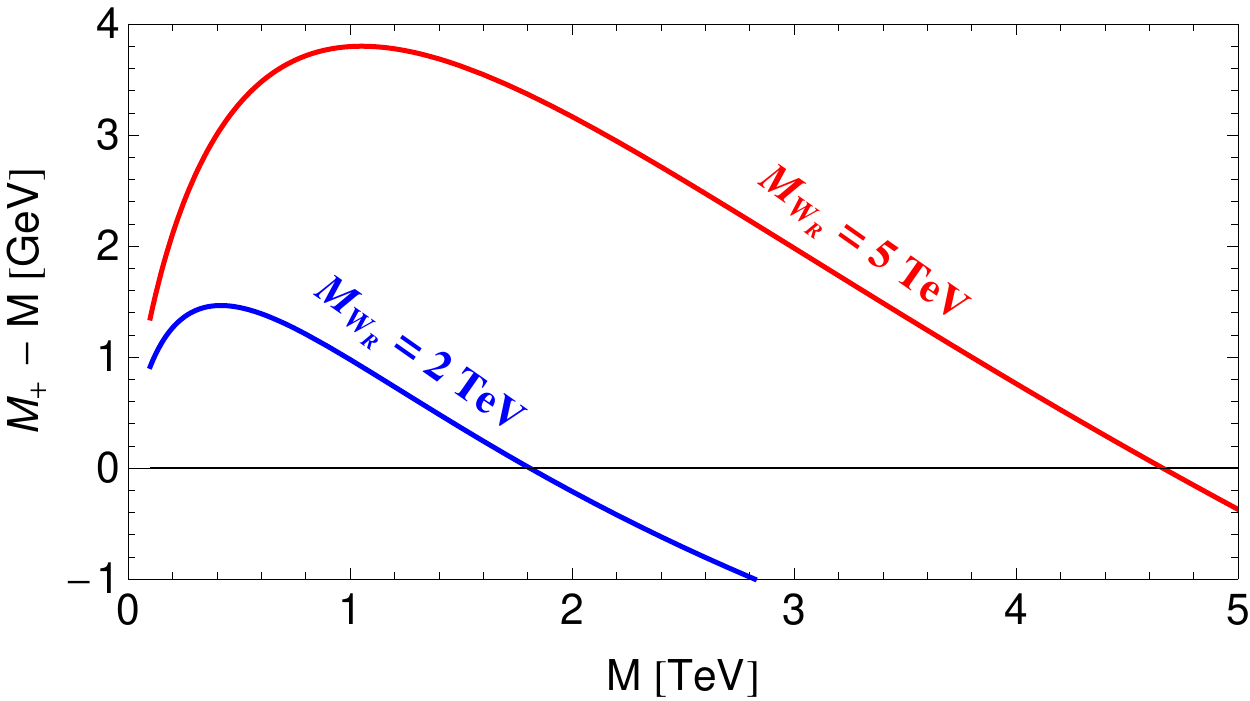}
\caption{Mass splitting for the right-handed triplet $\phi_R$.}
\label{fig:mass_splitting}
\end{figure}

The masses within each multiplet $\Psi_X$ will be split by radiative corrections. For the left-handed multiplet this is well known and increases the mass of the charged components by $\unit[166]{MeV}\,Q^2$, so the lightest component of $\Psi_L$ is neutral~\cite{Feng:1999fu}. 
Neglecting the small LR gauge boson mixing angles, we find the mass-splitting formula for the right-handed multiplet $\Psi_R$
\begin{align}
\begin{split}
M_Q - M_0 &\simeq \frac{\alpha_2 M}{4\pi} Q^2 \left[ f(r_{W_R}) - c_M^2 f(r_{Z_2})\right. \\
&\qquad \left.- s_W^2 s_M^2 f(r_{Z_1}) - s_W^2 f(r_\gamma)\right] ,
\end{split}
\end{align}
where $s_M = \sin \theta_M \equiv \tan \theta_W$~\cite{Duka:1999uc}, $r_X = M_X/M$, and
\begin{align}
f(r) \equiv 2 \int_0^1 \dd x \, (1+x) \log\left[ x^2 + (1-x) r^2\right] .
\end{align}
The main difference to the left-handed mass splitting is the different ratio of $M_{Z_2}/M_{W_R}\simeq 1.7$ due to the dominant triplet vacuum expectation value -- compared to the SM doublet case $M_{Z_1}/M_{W_L}\simeq 1.14$. 
This drives the mass splitting negative for $M\gtrsim 0.9\, M_{W_R}$ (see Fig.~\ref{fig:mass_splitting}), which would lead to unwanted charged DM. In a model with $M_{Z_2}/M_{W_R} =1.5$, the valid region would be extended to $M\lesssim 3.5\, M_{W_R}$, so the gauge boson mass ratio has a huge impact on the valid DM parameter space.

\subsection{Relic density}

Extending the LR model implementation for \texttt{CalcHep} of Ref.~\cite{Roitgrund:2014zka} by our new particles and interactions, we can evaluate the relic density $\Omega_R$ of $\Psi_R^0$ numerically using \texttt{micrOMEGAs}~\cite{Belanger:2006is,Belanger:2013oya}. 
The left-handed abundance $\Omega_L$ is more difficult to calculate because of strong Sommerfeld enhancement~\cite{Hisano:2006nn}; we will merely show the results from Refs.~\cite{Cirelli:2007xd,Cohen:2013ama}, which should still be valid in our case provided the mixing between the left and right gauge bosons is small.
The final abundance is then simply the sum
\begin{align}
\Omega h^2 = \Omega_L h^2 + \Omega_R h^2\,,
\end{align}
to be compared to the value measured by Planck~\cite{Ade:2013zuv}: $\Omega_\text{obs} h^2 = 0.1199\pm 0.0027$.

\begin{figure*}[t]
\includegraphics[width=0.45\textwidth]{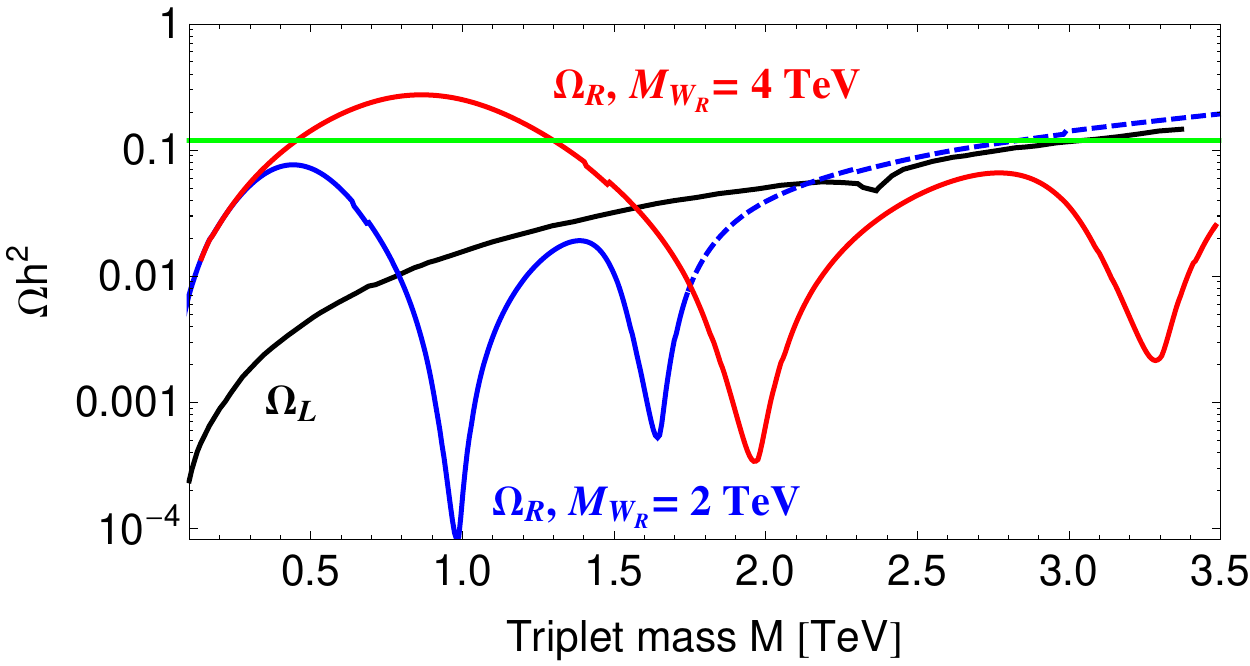}
\includegraphics[width=0.45\textwidth]{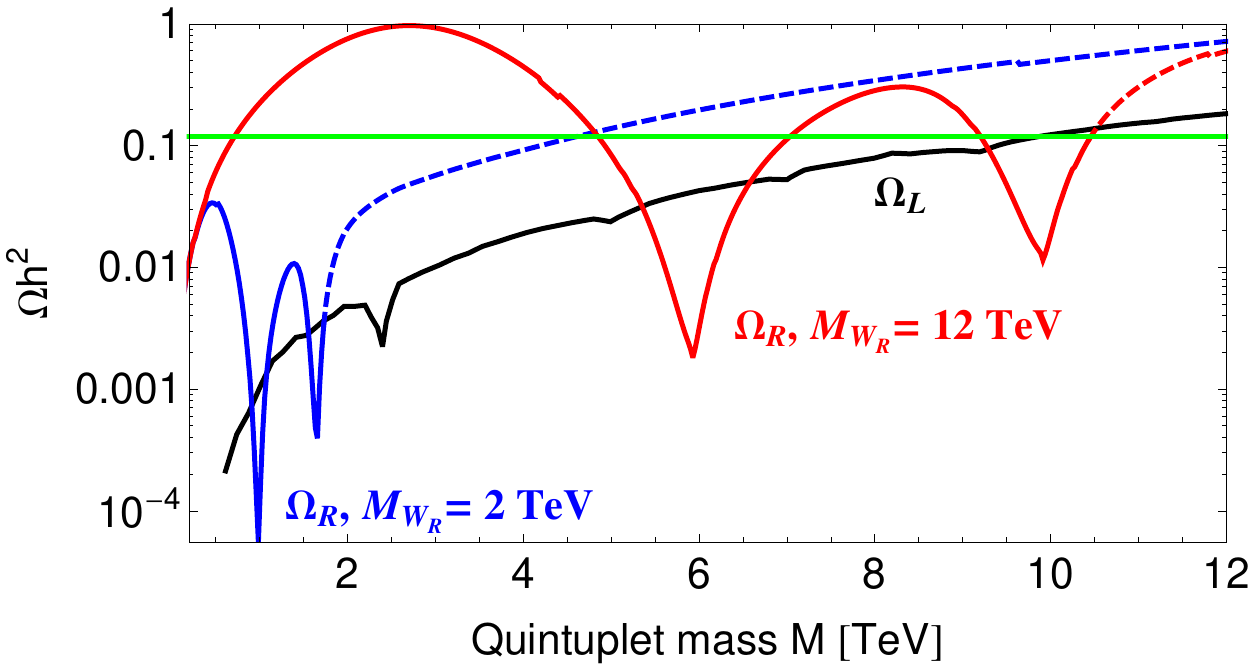}
\caption{Relic densities $\Omega_{L,R}$ for the LR fermion triplets (left) and quintuplets (right). $\Omega_L$ (black) includes the non-perturbative Sommerfeld enhancement and is taken from Refs.~\cite{Cirelli:2007xd,Cohen:2013ama}. $\Omega_R$ is shown for various $M_{W_R}$ in red and blue; clearly visible are the $W_R$ and $Z_R$ resonances. The dashed part is invalid for the taken relation $M_{Z_R} = 1.7\, M_{W_R}$ because DM is electrically charged, but a smaller $M_{Z_R}/M_{W_R}$ can make it valid.
The horizontal green line corresponds to the measured value.}
\label{fig:multiplets}
\end{figure*}

The results are shown for $n=1$ (wino-like triplet) and for $n=2$ (quintuplet) in Fig.~\ref{fig:multiplets} for various values of $M_{W_R}$. (Higher values of $n$ will lower the Landau pole of the theory, but can be studied in the same way.)
The left-handed abundance $\Omega_L$ provides an upper bound of $M\lesssim \unit[3]{TeV}$ ($\unit[10]{TeV}$) for the fermion triplet (quintuplet); clearly visible are the structures that arise due to Sommerfeld enhancement~\cite{Hisano:2006nn}. Additional resonances due to gauge boson mixing from $W_R$ and $Z_R$ would appear in a full analysis. While this can have some quantitative impact, it does not change the overall picture. (A resonance at the right place could for example allow for a heavier $M$ than allowed in standard MDM.) Let us instead focus on the right-handed abundance to see the possible effects.

\begin{figure}[t]
\includegraphics[width=0.35\textwidth]{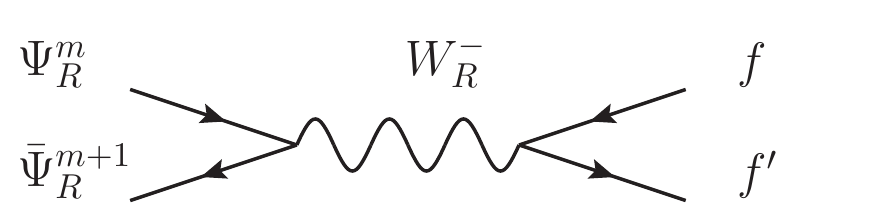}
\caption{Co-annihilation channels $\Psi_R^m \bar\Psi_R^{m+1}\to f f'$. The rotated diagram is relevant for the scattering $\Psi_R^0 f \leftrightarrow \Psi_R^+ f'$.}
\label{fig:coannihilations}
\end{figure}

For $M \lesssim M_{W_R}/4$, the relevant annihilation channel is $\Psi^{+,++}_R \Psi^{-,--}_R \to \,$SM; for $M \gtrsim M_{W_R}/4$, the depletion channel through $s$-channel co-annihilation via the $W_R$ opens up (Fig.~\ref{fig:coannihilations}), clearly visible in Fig.~\ref{fig:multiplets} at the resonance $M\sim M_{W_R}/2$. Around $ M \sim M_{Z_R}/2$ the $s$-channel $Z_R$ resonance appears, which depletes the charged components of $\Psi_R$. Since the scattering $\Psi_R^0 f \leftrightarrow \Psi_R^+ f'$ (Fig.~\ref{fig:coannihilations}) is fast in this region, it effectively also depletes the neutral component. 
In the region $M \gtrsim 0.9 \, M_{W_R}$ one would have charged DM because the mass splitting is negative, so it is not valid (dashed curves in Fig.~\ref{fig:multiplets}). However, in models with $M_{Z_R}/M_{W_R} < 1.7$ one can consider larger $M$ and eventually even reach a region $M >  M_{Z_R}/\alpha_R$ where Sommerfeld enhancement will be important.

\begin{figure}[b]
\includegraphics[width=0.45\textwidth]{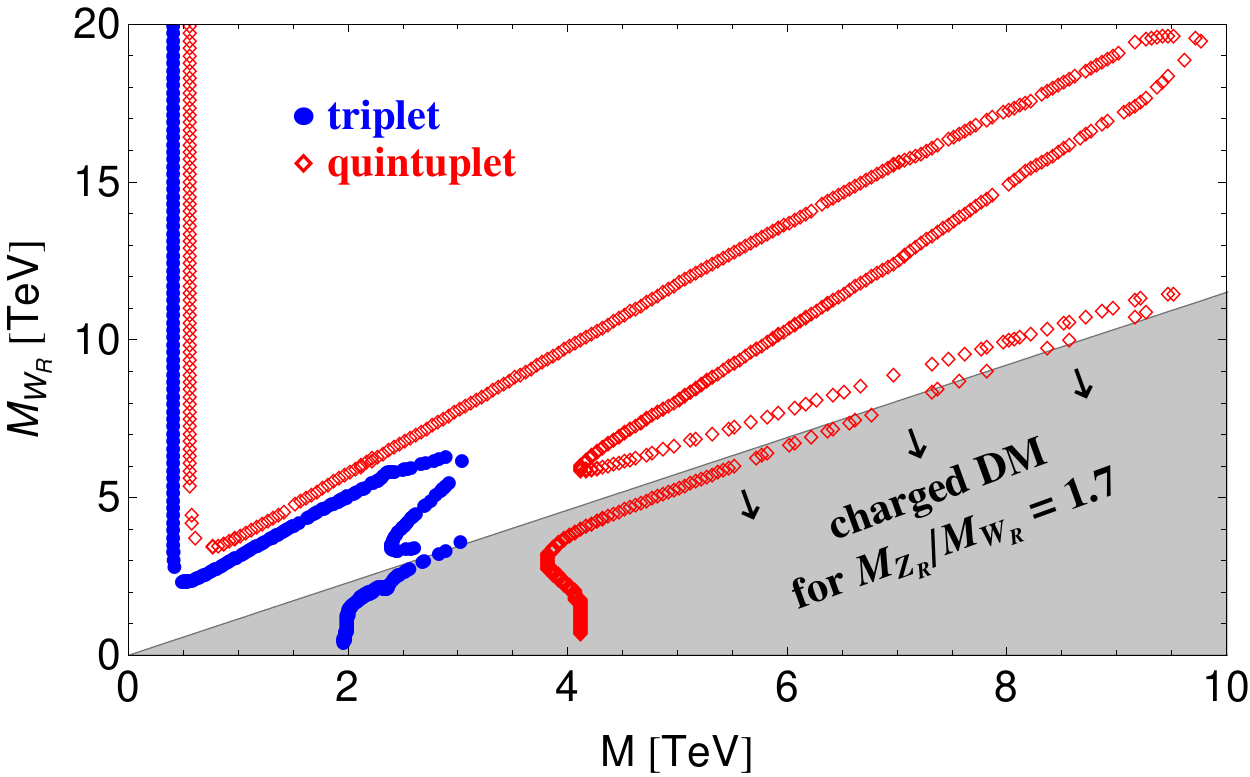}
\caption{Valid relic density $\Omega_L + \Omega_R = \Omega_\text{obs}$ for fermion triplets (blue dots) and quintuplets (red diamonds). In the gray region the mass splitting is negative, so DM is electrically charged; this region would be allowed for $M_{Z_R}/M_{W_R}<1.7$.}
\label{fig:MvsMWR}
\end{figure}

Overall, we see from Fig.~\ref{fig:MvsMWR} that one can easily realize the measured relic density in this minimal (two-component) LR DM scenario with $M_{Z_R}/M_{W_R} = 1.7$, provided $M_{W_R} \geq\unit[2.35]{TeV}$ for the triplet case ($M_{W_R} \geq\unit[3.43]{TeV}$ for quintuplets). Lower values for $M_{W_R}$ are possible for models with $M_{Z_R}/M_{W_R}<1.7$, which look qualitatively similar. 
Additional constraints on $M_{W_R}$ arise from collider searches~\cite{Khachatryan:2014dka}, which are however more model dependent and thus not shown in Fig.~\ref{fig:MvsMWR}.

DM detection signatures will be similar to the MDM case (see e.g.~Ref.~\cite{Cirelli:2007xd}), and one always has the subcomponent of left-handed MDM for which limits already exist. Loop-induced scattering of MDM off nucleons is rare~\cite{Hisano:2011cs,DelNobile:2013sia} and will not increase with the heavier gauge bosons of LR DM. Indirect detection via $\Psi_L^0 \Psi_L^0 \to \gamma \gamma$, $\gamma Z$, $W W$ is more promising and H.E.S.S.~already excludes $M\gtrsim \unit[1.6]{TeV}$ for triplet MDM with an Einasto profile (much weaker bounds possible for other profiles)~\cite{Cohen:2013ama}. A similar analysis will constrain the quintuplet~\cite{Aoki:2015nza}, but of course these bounds weaken if MDM is only a subcomponent of DM.
The LHC implications of LR dark matter are particularly interesting in the case $M < M_{W_R}/2$, as they modify the $W_R$ decay and allow for DM studies at the LHC if $W_R$ is observed. All of this will be explored in a future study~\cite{preparation}.

\subsection{Diboson excess}

Lastly, let us comment on the tantalizing diboson excess recently observed in ATLAS~\cite{Aad:2015owa}, which has spawned 
many explanations in terms of LR-related models~\cite{Cheung:2015nha,Dobrescu:2015qna,Aguilar-Saavedra:2015rna,Gao:2015irw,Brehmer:2015cia,Cao:2015lia}.\footnote{Ref.~\cite{Brehmer:2015cia} also commented on possible DM implications of the diboson solution, but very different from our full-model framework.}
In short, a new $W'$ gauge boson with mass $M_{W'}\sim \unit[2]{TeV}$ and couplings to quarks $g_R\sim 0.4$--$0.6$ can explain the resonance 
observed in the $W'\to W Z\to \,$jets channel, with signals in $ZZ$ and $WW$ due to jet misconstructions.\footnote{The prediction of 
$g_R$ close to $0.4$--$0.5$ by embedding the LR model into an $SO(10)$ have been carried out recently in Refs.~\cite{Deppisch:2014qpa,Heikinheimo:2014tba,Deppisch:2014zta,Patra:2015bga}.}
A potentially compatible small excess slightly below \unit[2]{TeV} has also been seen in CMS~\cite{Khachatryan:2014hpa,Khachatryan:2014gha}.

Any of the $SU(2)_1\times SU(2)_2\times U(1)'$ models aimed at explaining the diboson excess can be extended to include DM in the way outlined above. From Fig.~\ref{fig:MvsMWR} we see however that one needs to deviate from the relation $M_{Z_R}\simeq 1.7\, M_{W_R}$ in order to reach the measured abundance for $M_{W_R}= \unit[2]{TeV}$. This is possible in more general models and then allows for triplet DM with $M\simeq \unit[2.3]{TeV}$ or quintuplet DM with $M \simeq \unit[4.2]{TeV}$. The triplet case is actually already excluded by H.E.S.S.~for all realistic DM profiles, because the mass falls exactly in the Sommerfeld enhanced region~\cite{Cohen:2013ama}. We are hence drawn to quintuplet DM, dominantly consisting of $\Psi_R$. Since the gauge couplings in such a generalized model deviate from $g_R = g_L$, a model-dependent quantitative study is required to calculate the precise value of $M$ and ensure a positive mass splitting.

A more exciting possibility arises when we consider the introduction of two or three generations of DM multiplets. For example, two $\phi_R$ triplets with mass around $M\sim \unit[400]{GeV}$ can give a valid relic density for $M_{W_R}=\unit[2]{TeV}$. Since they are lighter 
than the $W_R$, the decay channels $W_R^+\to \Psi^0_R \Psi^+_R$ open up, together yielding a large branching ratio 
of $40$--$50\%$ (assuming $M_{\nu_R}\gg M_{W_R}$). This in effect weakens the bounds on $g_R$ and allows for a true LR value $g_R = g_L$, 
as mentioned in Refs.~\cite{Dobrescu:2015qna,Brehmer:2015cia}. Not only do we then solve the DM problem, but the resolution 
of the diboson excess can be a step closer to a LR model. 

With neutrino masses and dark matter taken care of, let us briefly mention the profound 
implications of a $\unit[2]{TeV}$ $W_R$ gauge boson on leptogenesis (for details, see 
Refs.~\cite{Carlier:1999ac,Frere:2008ct,Deppisch:2013jxa,Dev:2014iva,Dhuria:2015cfa}). 
The observed baryon asymmetry of the universe could be explained via leptogenesis where 
a net lepton asymmetry is generated due to out-of-equilibrium decay of the heavy Majorana 
neutrinos. It has recently been noted~\cite{Dhuria:2015cfa} that the scattering processes $W_{R}^{+} \ell_{R}^{-} \rightarrow W_{R}^{-} \ell_{R}^{+}$ through the doubly-charged scalar $\chi^{++}$ and heavy neutrinos $\nu_{R}$ -- having few TeV masses -- 
can be large enough to wash out any pre-existing lepton asymmetry. However, the lower bound on 
$M_{W_R}$ can be $9.9$ TeV with $g_L=g_R$ if flavored resonant leptogenesis is
taken into account \cite{Dev:2015vra}. A fully flavor resonant leptogenesis might allow for a
$2\,$TeV $W_R$ gauge boson consistent with the diboson excess and other low scale constraints if one can 
consider smaller $g_R$ values (e.g, $g_R/g_L\simeq 0.6$~\cite{Deppisch:2014qpa,Heikinheimo:2014tba,Deppisch:2014zta,Patra:2015bga}),
as the prominent scattering process $W_{R}^{+} \ell_{R}^{-} \rightarrow W_{R}^{-} \ell_{R}^{+}$ 
is now suppressed by at least a factor of $(g_R/g_L)^4 \simeq 8$. However, the detailed discussion is beyond the scope of this letter.

\section{Conclusion}

Left--right symmetric models based on the gauge group $SU(2)_L\times SU(2)_R\times U(1)_{B-L}$ are theoretically appealing because they restore parity at high energies and naturally provide small neutrino masses. We have presented a simple approach to also obtain stable DM in these models. Our framework is very minimal, with only one additional parameter -- the DM mass -- and no need for an ad-hoc stabilizing symmetry. New fermion (boson) multiplets can be stable either because they are even (odd) under 
the remaining unbroken subgroup $\mathbb{Z}_2^{B-L}$, or because they are of high enough $SU(2)$ dimension that 
they can not decay using dimension-4 operators. The simplest examples are fermion 
multiplets $(\vec{2 n+1},\vec{1},0) \oplus (\vec{1},\vec{2 n+1},0)$ with degenerate mass $M$ due to LR exchange 
symmetry. The left-handed component then acts as standard MDM, while the right-handed multiplet 
can have a different phenomenology due to the unknown $W_R$ and $Z_R$ masses. The observed relic density can be 
obtained in a variety of ways with TeV-scale DM, with plenty of signatures to be explored in future work.

The recent diboson excess in ATLAS, as well as a handful other excesses around the same mass in various channels, 
has spawned many explanations in terms of LR models with $M_{W_R}\simeq \unit[2]{TeV}$. All of them 
can potentially be extended to solve the DM issue in the way outlined here. The observation of a $2$~TeV $W_R$ gauge 
boson implies low scale $B-L$ violation which could be the first direct evidence for baryon and lepton number violation 
in nature and can have a strong implications on the generation of neutrino masses and the observed baryon asymmetry of the Universe.

\section*{Acknowledgements}

JH is very grateful to Camilo Garcia-Cely and Michel Tytgat for enlightening discussions and comments on the manuscript.
The work of JH is funded in part by IISN and by Belgian Science Policy (IAP VII/37).
The work of SP is partially supported by the Department of Science and 
Technology, Govt.~of India under the financial grant SB/S2/HEP-011/2013 
and by the Max Planck Society in the project MANITOP.

\textbf{Note added:}
During the publication process of this letter new analyzes appeared that derive indirect detection constraints of the MDM quintuplet and hence also constrain the LR version discussed here~\cite{Cirelli:2015bda,Garcia-Cely:2015dda}.

\bibliographystyle{utcaps_mod}
\bibliography{BIB}

\end{document}